 \documentclass[aps,prd,twocolumn,showpacs,floatfix,nofootinbib,superscriptaddress]{revtex4}
\usepackage{dcolumn}

\usepackage{amsmath,amsfonts,extarrows,amssymb,mathrsfs,times,bm}
\usepackage{extarrows}
\usepackage{graphicx}
\usepackage{float}
\usepackage{graphicx,epstopdf}
\usepackage{dcolumn}
\usepackage{exscale}
\usepackage{relsize}
\usepackage{mathtools}
\usepackage{mhchem}
\usepackage[colorlinks=true,breaklinks=true,linkcolor=blue,citecolor=blue,urlcolor=blue]{hyperref}
\newcommand{\nn}{\nonumber}

\begin{document}

\title{Circular orbit of a particle and weak gravitational lensing}
\author{Zonghai Li}
\email{lzh@my.swjtu.edu.cn}
\affiliation{MOE Key Laboratory of Artificial Micro- and Nano-structures, School of Physics and Technology, Wuhan University, Wuhan 430072, China}
\date{\today}
\author{Guodong Zhang}
\email{yashangluzhu@snnu.edu.cn}
\affiliation{Chengguan Middle School, LuoNan,726100,China}
\author{Ali {\"O}vg{\"u}n}
\email{ali.ovgun@emu.edu.tr}
\affiliation{Physics Department, Arts and Sciences Faculty, Eastern Mediterranean
University, Famagusta, North Cyprus 99628 via Mersin 10, Turkey.}

\begin{abstract}
The purpose of this paper is twofold. First, we introduce a geometric approach to study the circular orbit of a particle in static and spherically symmetric spacetime based on Jacobi metric. Second, we apply the circular orbit to study the weak gravitational deflection of null and time-like particles based on Gauss-Bonnet theorem. By this way, we obtain an expression of deflection angle and extend the study of deflection angle to asymptotically non-flat black hole spacetimes. Some black holes as lens are considered such as a static and spherically symmetric black hole in the conformal Weyl gravity and a Schwarzschild-like black hole in bumblebee gravity. Our results are consistent with the previous literature. In particular, we find that the connection between Gaussian curvature and the radius of a circular orbit greatly simplifies the calculation.
\end{abstract}


\maketitle

\section{Introduction}
In differential geometry, curvature is a core concept. For Riemannian manifolds with dimensions greater than two, its information is described by the Riemann curvature tensor. But in a two-dimensional manifold, the situation is simpler because the Gaussian curvature (equivalently, scalar curvature) already contains all the information~\cite{Ruppeiner1995}. On the other hand, in general theory of relativity, gravity is not a force, but a curvature of spacetime. Naturally, one would hope that gravitational-related effects could be represented by curvature.

Light deflection is one of the predictions of general relativity, which has been verified by a series of experiments~\cite{DED1920,Will2015}. Nowadays the gravitational lensing becomes a important tool in astronomy and cosmology. For example, it is used to measure the mass of galaxies and clusters~\cite{Hoekstra2013,Brouwer2018,Bellagamba2019}, to detect dark matter and dark energy~\cite{Vanderveld2012,cao2012,zhanghe2017,Huterer2018,SC2019,Andrade2019,Turimov:2018ttf}. For a light ray propagating through the equatorial plane, how does the deflection angle  be represented by the Gauss curvature? In 2008, Gibbons and Werner applied the Gauss-Bonnet (GB) theorem to study the weak gravitational deflection angle of light in static and spherically symmetric (SSS) gravitational field~\cite{GW2008}. In their geometric method, the deflection angle can be calculated by integrating the Gaussian curvature of corresponding optical metric. The importance of the Gibbons-Werner method is that it shows that the deflection angle can be viewed as a global effect. How to extend this geometrical method to the stationary and axially symmetric (SAS) spacetimes?  For a stationary spacetime, the corresponding optical geometry is defined by a Randers-Finsler metric. Bloomer~\cite{Bloomer2011} tried early and was eventually established by Werner~\cite{Werner2012} using Naz{\i}m's osculating Riemannian manifold method~\cite{Nazim1936}.

With the Gibbons-Werner method, the weak gravitational deflection of light by different lens objects in differential gravity models have been widely studied in Refs~\cite{Jusufi:string17,Jusufi&Ali:Teo,Jusufi&Ali:string,Jusufi:RB,Jusufi:monopole,Ali:wormhole,Ali:strings,Ali:BML,Javed1,Javed2,Javed3,Javed4,Sakalli2017,Goulart2018,Leon2019,LZ20201,zhu2019,444,445,446,447,448,449,450}. Furthermore, some authors used GB theorem to study the deflection of massive particles ~\cite{CG2018,CGJ2019,Jusufi:mp,LHZ2020}. Recently a new step forward was put by Ishihara \textit{et al.}~\cite{ISOA2016,IOA2017} extending the use of the Gibbons-Werner method to more general situations where the receiver and source are assumed to be at finite distance from a lens. More, Ono \textit{et al.}~\cite{OIA2017,OIA2018,OIA2019,OA2019} introduced the generalized optical metric method to study the finite-distance deflection of light in SAS spacetimes. Very recently, by using Jacobi-Maupertuis Randers-Finsler metric and GB theorem, Li~\textit{et al.} studied the finite-distance effects on gravitational deflection of massive particles in SAS spacetimes~\cite{LA2020,LJ2020,LZ20202}.

Both in the case of finite-distance deflection and infinite-distance deflection, the GB theorem is often applied to an infinite region outside of the particle ray. However, it cannot usually be applied to study of deflection angle in some asymptotically non-flat spacetimes. Taking Schwarzschild-ds spacetime as an example, the terms containing the cosmological constant is divergent as radial coordinate r approaches infinity. Arakida~\cite{Arakida2018} first considered a finite region and study the deflection angle of light in Schwarzschild-ds spacetime. Very recently, Takizawa~\textit{et al.}~\cite{Takizawa2020} used non-geodesic circular orbits with a minimum distance as radius to form two finite regions and studied the deflection angle of light in asymptotically non-flat spacetimes. However, in order to calculate the deflection angle, the geodesic curvature of non-geodesic circular orbit needs to be considered in Ref.~\cite{Takizawa2020}. The idea of this article is to use a geodesic circular orbit instead of a non-geodesic circular orbit to avoid geodesic curvature terms in deflection angle. To this end, we will first introduce a geometric approach to derive the radius of circular orbit of particle in SSS spacetimes. Then, we will apply the circular orbit and GB theorem to study the weak deflection effects of particle. 

This paper is organized as follows. In Sec.~\ref{COPJ}, based on Jacobi metric we apply a geometric method to derive the radius of circular orbit of particle moving in the equatorial plane of SSS spacetimes. In Sec.~\ref{lensgeometry}, we apply the circular orbit of particle to study the weak gravitational deflection angle. In Sec.~\ref{examples}, we calculate the weak deflection angle of particle in two asymptotically non-flat spacetimes. Finally, we end our paper with a short conclusion in Sec.~\ref{conclusion}. We set $G = c = 1$ in this paper.

\section{Circular orbits of a particle in SSS spacetimes: a Jacobi metric method}\label{COPJ}
\subsection{Jacobi metric}
The Jacobi metric of curved spacetime is an important tool for studying gravitational effects. The Jacobi metric of static spacetime is established by Gibbons~\cite{Gibbons2016}, while the Jacobi metric of stationary spacetimes is established by Chanda~\textit{et al.}~\cite{Chanda2019,Chanda2019b}. In this subsection, we briefly review Jacobi metric of SSS spacetimes (see Refs.~\cite{Gibbons2016,LHZ2020} for details). For convenience, we use $g_{ij}$ to denote Jacobi metric and this is followed for the quantities with Jacobi metric. For a SSS metric,
\begin{eqnarray}
\label{SSS}
 ds^2&=&\bar{g}_{\mu\nu}dx^\mu dx^\nu\nn\\
 &=&-A(r)dt^2+B(r)dr^2+C(r)d\Omega^2,
\end{eqnarray}
its Jacobi metric reads
\begin{eqnarray}
 dl^2&=&g_{ij}dx^idx^j\nn\\
 &=&\left(E^2-m^2A\right)\left(\frac{B}{A}dr^2+\frac{C}{A}d\Omega^2\right),
\end{eqnarray}
where $m$ and $E$ denote the energy and mass of a particle, respectively, and $d\Omega^2=d\theta^2+\sin^2\theta d\phi^2$ is the line element of unit two-sphere.
The Jacobi metric reduces to optical metric as $m=0$ and $E=1$,
\begin{eqnarray}
\label{opticalmetric}
 &&dl^2=dt^2=\frac{B}{A}dr^2+\frac{C}{A}d\Omega^2.
\end{eqnarray}
The energy of particle at infinity for an asymptotic observer is
\begin{eqnarray}
 &&E=\frac{m}{\sqrt{1-v^2}},
\end{eqnarray}
where $v$ is the velocity of particle.
Without loss of generality, we study the motion of particles in the equatorial plane ($\theta = \pi/2 $). Then, the Jacobi metric becomes
\begin{eqnarray}
\label{Jacobi}
 &&dl^2=m^2\left(\frac{1}{1-v^2}-A\right)\left(\frac{B}{A}dr^2+\frac{C}{A}d\phi^2\right).
\end{eqnarray}
The orbit equation of a particle moving in equatorial plane can be written as~\cite{LHZ2020}
\begin{eqnarray}
\label{particleorbit}
 &&\left(\frac{du}{d\phi}\right)^2=\frac{C^4u^4}{AB}\left[\frac{1}{b^2v^2}-A\left(\frac{1-v^2}{b^2v^2}-\frac{1}{C}\right)\right],
\end{eqnarray}
where $u=1/r$ and $b$ is the impact parameter. Let $v=1$, it leads to the orbit equation of light as follows
\begin{eqnarray}
\label{lightorbit}
 &&\left(\frac{du}{d\phi}\right)^2=\frac{C^4u^4}{AB}\left[\frac{1}{b^2}-\frac{A}{C}\right].
\end{eqnarray}
\subsection{The circular orbit of a particle in equatorial plane }
Suppose the circular orbit $\gamma_{co}$ is defined by $r=r_{co}=constant$, then, by Eq.~\eqref{Jacobi} we have
\begin{eqnarray}
 &&dl^2=m^2\left[\frac{1}{(1-v^2)A(r_{co})}-1\right]C(r_{co})d\phi^2.
\end{eqnarray}
The geodesic curvature of curve $\gamma_{co}$ is~\cite{LHZ2020}
\begin{eqnarray}
\label{geodesicurvature}
 &&\kappa(\gamma_{co})=\sqrt{g_{rr}\left(\Gamma_{\phi\phi}^r\right)^2}\left(\frac{d\phi}{dl}\right)^2\mid_{r=r_{co}}.
\end{eqnarray}
Let $\kappa(\gamma_{co})=0$, by metric~\eqref{Jacobi}, the radius of circular orbits satisfy,
\begin{eqnarray}
\label{particleCO}
 0&=&C(r_{co})\partial_{r}A(r_c)-A(r_{co})\partial_{r}C(r_{co})\nn\\
 &&\times\left[1-A(r_{co})+v^2A(r_{co})\right]
\end{eqnarray}
One can see that this equation is independent in metric function $B(r)$. For light ($v=1$), one can get
\begin{eqnarray}
\label{lightco}
 &&C(r_{co})\partial_{r}A(r_{co})-A(r_{co})\partial_{r}C(r_{co})=0.
\end{eqnarray}
When $C(r)=r^2$, this leads to
\begin{eqnarray}
 &&r_{co}\partial_{r}A(r_{co})-2A(r_{co})=0,
\end{eqnarray}
which is consistent with Ref.~\cite{Wei2018}.

\subsection{Example: circular orbits of a particle in Schwarzschild spacetime}
For Schwarzschild black hole, one has,
\begin{eqnarray}
A(r)&=&1-\frac{2M}{r},\nn\\
B(r)&=&\left(1-\frac{2M}{r}\right)^{-1},\nn\\
C(r)&=&r^2.\nn
\end{eqnarray}
Here and below, we use $M$ to denote the mass of a certain black hole.
Substituting the above metric functions $A(r)$ and $C(r)$ into Eq.~\eqref{particleCO}, one can obtain the following circular orbit radius
\begin{eqnarray}
\label{SPCO}
r_{co}=\frac{\left(-1+4v^2+\sqrt{1+8v^2}\right)M}{2v^2}.
\end{eqnarray}
For $v=1$, it reduces to the radius of circular orbit of light
\begin{eqnarray}
\label{SLCO}
r_{co}=3M.
\end{eqnarray}
\section{Application circular orbit to weak gravitational lensing}\label{lensgeometry}
\subsection{Gauss-Bonnet theorem}
Let $D$ be a compact oriented two-dimensional Remannian manifold with Gaussian curvature $\mathcal{K}$ and Euler characteristic $\chi(D)$, and its boundary $\partial{D}$ is a piecewise smooth curve with geodesic curvature $k$. Then GB theorem states that~\cite{GW2008,Carmo1976}
\begin{equation}
\iint_D{\mathcal{K}}dS+\oint_{\partial{D}}k~d\sigma+\sum_{i=1}{\beta_i}=2\pi\chi(D),\\
\end{equation}
where $dS$ is the area element, $d\sigma$ is the line element of boundary, and $\beta_i$ is the jump angle in the $i$-th vertex of $\partial{D}$ in the positive sense, respectively.

\subsection{Lens geometry}
In this subsection, we will use the GB theorem to study the gravitational deflection of particle. We apply the definition of deflection angle proposed in Ref.~\cite{ISOA2016}
\begin{eqnarray}
\label{angle}
&& \alpha\equiv \Psi_R-\Psi_S+\phi_{RS},
\end{eqnarray}
where $\Psi_R$ and $\Psi_S$ are angles between the tangent of the particle ray and the radial direction from the lens to receiver and source, respectively, and the coordinate angle $\phi_{RS}\equiv\phi_R-\phi_S$.
\begin{figure}[t]
\label{Figure}
\centering
\includegraphics[width=8.0cm]{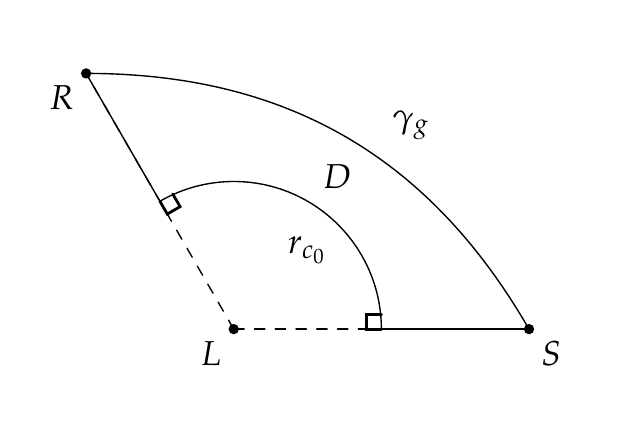}
\caption{A region $D\subset(M,g_{ij})$. Notice that $\beta_R=\pi-\Psi_R$ and $\beta_S=\Psi_S$.}
\end{figure}

Now, we consider a region $D\subset(M,g_{ij})$ bounded by four geodesics: a particle ray $\gamma_{g}$ from source ($S$) to receiver ($R$), particle circular orbit $\gamma_{co}$, ,two spatial geodesics of outgoing radial lines passing through $R$ and $S$ respectively. Due to the fact that $D$ is a non singular region, one can see $\chi(D_{r_0})=1$. In addition, the circular orbit $\gamma_{co}$ is perpendicular to the radial lines. Considering the above facts, applying GB theorem to region $D$, we can get
\begin{eqnarray}
&&\iint_{D}\mathcal{K} dS+\beta_S+\beta_R=\pi.
\end{eqnarray}
Using the definition~\eqref{angle}, considering $\beta_R=\pi-\Psi_R$ and $\beta_S=\Psi_S$, we can obtain the deflection angle as follows
\begin{eqnarray}
\label{INGBangle}
\alpha&=&\iint_{D}\mathcal{K} dS+\phi_{RS}.
\end{eqnarray}
The expression of deflection angle~\eqref{INGBangle} is applicable to asymptotically non-flat spacetime, and it also applies to study the finite-distance effects. More, it is more concise than in Ref.~\cite{Takizawa2020}. However, it should be noted that the formula~\eqref{INGBangle} is only applicable to the case where circular orbits of particle exists in spacetime. Finally, the Gaussian curvature of Jacobi metric can be calculated as~\cite{Werner2012}
\begin{eqnarray}
\label{Gauss-K}
\mathcal{K}&=&\frac{1}{\sqrt{\det g}}\left[\frac{\partial \left(\frac{\sqrt{\det g}}{g_{rr}}{{\Gamma}^\phi_{rr}}\right)}{\partial{\phi}}-\frac{\partial\left(\frac{\sqrt{\det g}}{g_{rr}}{{\Gamma}^\phi_{r\phi}}\right)}{\partial{r}}\right].~~~~~
\end{eqnarray}

\subsection{The integration of Gaussian curvature}
 From optical metric~\eqref{opticalmetric}, we can get
\begin{eqnarray}
&&g_{rr}=\frac{B}{A},\nn\\
&&detg=\frac{BC}{A^2},\nn\\
&&\Gamma_{r\phi}^{\phi}=\frac{1}{2}\left(\frac{C'}{C}-\frac{A'}{A}\right),\nn
\end{eqnarray}
where $'$ denotes derivative with respect to $r$.

Then, by Eq.~\eqref{Gauss-K}, we have
\begin{eqnarray}
\int\mathcal{K}\sqrt{detg}dr&=&-\int\frac{d\left(\frac{\sqrt{\det g}}{g_{rr}}{{\Gamma}^\phi_{r\phi}}\right)}{d{r}}dr,\nn\\
&=&-\frac{\sqrt{\det g}}{g_{rr}}{{\Gamma}^\phi_{r\phi}}\nn\\
&=&\frac{CA'-AC'}{2A\sqrt{BC}}.
\end{eqnarray}
Note that we ignore the constant of integration here, which has no effect on the study in this article. Then, according to Eq.~\eqref{lightco}, we have
 \begin{eqnarray}
 \label{coGauss}
\left[\int \mathcal{K}\sqrt {det g}dr\right]_{r=r_{co}}=0.
\end{eqnarray}
Although we only discuss optical metrics, it is not hard to believe that this property holds true for the Jacobi metric. Eq.~\eqref{coGauss} can greatly simplify the calculation of the deflection angle, because it can reduce Eq.~\eqref{INGBangle} to
\begin{eqnarray}
\label{SINGBangle}
\alpha &=&\int_{\phi_{S}}^{\phi_{R}}\int_{r_{co}}^{r(\phi)} \mathcal{K}\sqrt{detg} drd\phi+\phi_{RS}\nn\\
&=&\int_{\phi_{S}}^{\phi_{R}}\left[\int \mathcal{K}\sqrt{detg} dr\right]_{r=r(\phi)}d\phi+\phi_{RS}.
\end{eqnarray}
In the next section, we will use Eq.~\eqref{INGBangle} to calculate the deflection angles of light and massive particles in asymptotically non-flat spacetimes.

\section{EXAMPLES IN ASYMPTOTICALLY NON-FLAT SPACETIMES}\label{examples}
In this section, we give two examples as applications of Eq.~\eqref{INGBangle}. In the first example, we study the deflection angle of light in a SSS spacetime in Weyl gravity model, and we need to use optical metrics as the background space. In the second example, we study the deflection angle of particles in Schwarzschild-like spacetime in bumblebee gravity model, using the Jacobi metric as the background space.
\subsection{Deflection angle of light in a static and spherically symmetric spacetime in the conformal Weyl gravity}
The line element of a static and spherically symmetric spacetime in the conformal Weyl gravity reads~\cite{Mannheim1989}
\begin{eqnarray}
\label{Weyl}
ds^2&=&-\left(1-3M\gamma-\frac{2M}{r}+\gamma r-kr^2\right)dt^2\nn\\
&&+\left(1-3M\gamma-\frac{2M}{r}+\gamma r-kr^2\right)^{-1}dr^2\nn\\
&&+r^2\left(d\theta^2+\sin^2\theta d\phi^2\right),
\end{eqnarray}
where $\gamma$ and $k$ are the constants. For simplify, we choose $k=0$. Substituting metric~\eqref{Weyl} into Eq.~\eqref{lightco}, one can obtain the radius of circular orbits of light as follows,
\begin{eqnarray}
\label{RNR}
r_{co}&=&\frac{-1+3M\gamma+\sqrt{1+9M^2\gamma^2}}{\gamma}.
\end{eqnarray}
Using Eq.~\eqref{lightorbit}, we can obtain the orbit equation of light in this spacetime
\begin{eqnarray}
\left(\frac{du}{d\phi}\right)^2&=&\frac{1}{b^2}-u^2+2Mu^3-u\gamma+3Mu^2\gamma.
\end{eqnarray}
We can get its solution by iterative method as follows
\begin{eqnarray}
\label{Weylorbit}
u&=&\frac{\sin\phi}{b}+\frac{\left(1+\cos^2\phi\right)M}{b^2}-\frac{\gamma}{2}+\mathcal{O}\left(M^2,\gamma^2\right),
\end{eqnarray}
which can derive the coordinate angles as
\begin{eqnarray}
\label{WeySangle}
\phi_S&=&\arcsin(bu_S)-\frac{\left(2-b^2u_S^2\right)M}{b\sqrt{1-b^2u_S^2}}+\frac{b\gamma}{2\sqrt{1-b^2u_S^2}}\nn\\
&&-\frac{b^3u_S^3\gamma M}{2\left(1-b^2u_S^2\right)^{3/2}}\lambda+\mathcal{O}\left(M^2,\gamma^2\right),\\
\label{WeyRangle}
\phi_R&=&\pi-\arcsin(bu_R)+\frac{\left(2-b^2u_R^2\right)M}{b\sqrt{1-b^2u_R^2}}\nn\\
&&-\frac{b\gamma}{2\sqrt{1-b^2u_R^2}}+\frac{b^3u_R^3\gamma M}{2\left(1-b^2u_R^2\right)^{3/2}}\lambda\nn\\
&&+\mathcal{O}\left(M^2,\gamma^2\right),
\end{eqnarray}
In addition, the corresponding optical metric of metric~\eqref{Weyl} is
\begin{eqnarray}
\label{Kottlermetric}
dt^2&=&\frac{dr^2}{\left(1-3M\gamma-\frac{2M}{r}+\gamma r\right)^{2}}\nn\\
&&+\frac{r^2 d\phi^2 }{1-3M\gamma-\frac{2M}{r}+\gamma r},
\end{eqnarray}
with Gaussian curvature
\begin{eqnarray}
\label{KottlerGauss}
\mathcal{K}&=&-\frac{\gamma^2}{4}-\frac{\left(2+3r\gamma\right)M}{r^3}+\frac{3\left(1+2r\gamma\right)M^2}{r^4}.
\end{eqnarray}

Thus, we have
\begin{eqnarray}
&&\iint_D\mathcal{K}dS=\int_{\phi_S}^{\phi_R}\int_{r_{co}}^{r(\phi)}\mathcal{K}\sqrt{detg}dr d\phi\nn\\
&&=\int_{\phi_S}^{\phi_R}\left[\frac{6M\left(1+r\gamma\right)-r\left(2+r\gamma\right)}{2r\sqrt{1-3M\gamma-\frac{2M}{r}+\gamma r}}\right]_{r=r(\phi)}d\phi\nn\\
&&=\int_{\phi_S}^{\phi_R}\left(-1+\frac{2M\sin\phi}{b}+\mathcal{O}\left(M^2,\gamma^2\right)\right)d\phi\nn\\
&&=\frac{2M\left(\cos \phi_S-\cos \phi_R\right)}{b}-\phi_{RS}+\mathcal{O}\left(M^2,\gamma^2\right),~~~~~
\end{eqnarray}
where we used Eq.~\eqref{Weylorbit} and $r=1/u$.

Finally, the deflection angle becomes
\begin{eqnarray}
\alpha&=&\iint_{D}\mathcal{K} dS+\phi_{RS}\nn\\
&=&\frac{2M\left(\cos \phi_S-\cos \phi_R\right)}{b}+\mathcal{O}\left(M^2,\gamma^2\right)\nn\\
&=&\frac{2M\left(\sqrt{1-b^2u_R^2}+\sqrt{1-b^2u_S^2}\right)}{b}\nn\\
&&-Mb\gamma\left(\frac{u_R}{\sqrt{1-b^2u_R^2}}+\frac{u_S}{\sqrt{1-b^2u_S^2}}\right)\nn\\
&&+\mathcal{O}\left(M^2,\gamma^2\right),
\end{eqnarray}
where we used Eqs.~\eqref{WeySangle} and~\eqref{WeyRangle}. This result is the same as that calculated by $\alpha\equiv \Psi_R-\Psi_S+\phi_{RS}$ in Ref.~\cite{ISOA2016}, and calculated by $\alpha=\iint_{D_R+D_S}\mathcal{K}dS+\int_{P_R}^{P_S}\kappa_gdl+\phi_{RS}$ in Ref.~\cite{Takizawa2020}.

\subsection{Deflection angle of massive particle in Schwarzschild-like spacetime in bumblebee gravity}
The Schwarzschild-like black hole in bumblebee gravity model is given as~\cite{Casana2018}
\begin{eqnarray}
\label{SLB}
ds^2&=&-\left(1-\frac{2M}{r}\right)dt^2+\frac{\lambda^2}{\left(1-\frac{2M}{r}\right)}dr^2\nn\\
&&+r^2\left(d\theta^2+\sin^2\theta d\phi^2\right),
\end{eqnarray}
where $\lambda=\sqrt{1+l}$ with $l$ being the Lorentz violation parameter. From this, one can see that $\lambda$ is only included in the metric function $B(r)$. Therefore, the particle circular orbit is the same as in the case of Schwarzschild black hole,
\begin{eqnarray}
\label{SLPCO}
r_{co}=\frac{\left(-1+4v^2+\sqrt{1+8v^2}\right)M}{2v^2}.
\end{eqnarray}

The orbit equation is
\begin{eqnarray}
\left(\frac{du}{d\phi}\right)^2&=&\frac{2Mu\left(1-v^2+b^2u^2v^2\right)}{b^2v^2\lambda^2}\nn\\
&&+\frac{u^2}{\lambda^2}\left(\frac{1}{b^2}-u^2\right)+\mathcal{O}\left(M^2\right),
\end{eqnarray}
with its iterative solution
\begin{eqnarray}
\label{Slikeorbit}
u&=&\frac{\sin(\frac{\phi}{\lambda})}{b}+\frac{\left(1+v^2\cos^2(\frac{\phi}{\lambda})\right)M}{b^2v^2}+\mathcal{O}\left(M^2\right).
\end{eqnarray}
As well, we have
\begin{eqnarray}
\label{angleS}
\phi_S&=&-\frac{\left(1+v^2-b^2u_S^2v^2\right)M\lambda}{b\sqrt{1-b^2u_S^2}v^2}\nn\\
&&+\lambda \arcsin(bu_S)+\mathcal{O}\left(M^2\right),\\
\label{angleR}
\phi_R&=&\frac{\left(1+v^2-b^2u_R^2v^2\right)M\lambda}{b\sqrt{1-b^2u_R^2}v^2}\nn\\
&&+\lambda \left(\pi-\arcsin(bu_R)\right)+\mathcal{O}\left(M^2\right)\\
\label{angleRS}
\phi_{RS}&=&\phi_R-\phi_S\nn\\
&=&\lambda\left[\pi-\arcsin(bu_R)-\arcsin(bu_S)\right]\nn\\
&&+\frac{M\lambda}{bv^2}\left(\frac{1}{\sqrt{1-b^2u_R^2}}+\frac{1}{\sqrt{1-b^2u_S^2}}\right)\nn\\
&&+\frac{M\lambda}{b}\left(\sqrt{1-b^2u_R^2}+\sqrt{1-b^2u_S^2}\right).
\end{eqnarray}
Now, we write the corresponding Jacobi metric as follows
\begin{eqnarray}
\label{JacobiS}
dl^2&=&m^2\left(\frac{v^2}{1-v^2}+\frac{2M}{r}\right)\nn\\
&&\times\left[\frac{\lambda^2dr^2}{\left(1-\frac{2M}{r}\right)^2}+\frac{r^2d\phi^2}{1-\frac{2M}{r}}\right],~~~~~
\end{eqnarray}

One can obtain its Gaussian curvature by Eq.~\eqref{Gauss-K},
\begin{eqnarray}
\label{JacobiS}
\mathcal{K}&=&\frac{M(1-v^2)}{m^2r^3\left[rv^2+2M(1-v^2)\right]^3\lambda^2}\nn\\
&&\times \bigg[8M^3(1-v^2)^2-r^3v^2(1+v^2)\nn\\
&&-3M r^2v^2(1-2v^2)\nn\\
&&-6M^2r(1-3v^2+2v^4)\bigg].
\end{eqnarray}

Then, one can obtain the integral of Gaussian curvature as follows
\begin{eqnarray}
\label{Kangle}
\iint_D\mathcal{K}dS&=&\frac{\left(1+v^2\right)M[\cos(\frac{\phi_S}{\lambda})-\cos(\frac{\phi_R}{\lambda})]}{bv^2}\nn\\
&&-\frac{\phi_{RS}}{\lambda}+\mathcal{O}\left(M^2\right),~~~~
\end{eqnarray}
where we used Eq.~\eqref{Slikeorbit}.
Finally, the deflection angle is
\begin{eqnarray}
\label{Sfangle}
\alpha&=&\iint_{D}\mathcal{K}  dS+\phi_{RS}\nn\\
&=&\frac{\left(1+v^2\right)M\left[\cos(\frac{\phi_S}{\lambda})-\cos(\frac{\phi_R}{\lambda})\right]}{bv^2}\nn\\
&&+\left(1-\frac{1}{\lambda}\right)\phi_{RS}+\mathcal{O}\left(M^2\right)\nn\\
 &=&\left(\lambda-1\right)\left[\pi-\arcsin(b u_R)-\arcsin(b u_S)\right]\nn\\
 &&+\bigg[\frac{\left(1+v^2\right)\lambda-b^2 u_R^2\left(1+v^2\lambda\right)}{\sqrt{1-b^2u_R^2}}\nn\\
 &&+\frac{\left(1+v^2\right)\lambda-b^2 u_S^2\left(1+v^2\lambda\right)}{\sqrt{1-b^2u_S^2}}\bigg]\frac{M}{bv^2}\nn\\
 &&+\mathcal{O}(M^2)~,
\end{eqnarray}
where we used Eqs.~\eqref{angleS},~\eqref{angleR} and~\eqref{angleRS}.
As well, let $u_S\rightarrow0$, and $u_R\rightarrow0$ the infinite-distance deflection angle can be obtained as
{\begin{eqnarray}
\label{Sinangle}
 \alpha_{\infty}
 &=&\left(\lambda-1\right)\pi+\frac{2\left(1+v^2\right)\lambda M}{bv^2}+\mathcal{O}(M^2)~.
\end{eqnarray}}
Eqs.~\eqref{Sfangle} and~\eqref{Sinangle} agree with that calculated by $\alpha=\iint_{_{R}^{\infty}\Box_{S}^{\infty}}\mathcal{K}dS+\int_{S}^{R}k_gd\sigma+\left(1-\frac{1}{\lambda}\right)\phi_{RS}$ in Ref.~\cite{LA2020}.
\section{conclusion} \label{conclusion}

In this paper, we have introduced a geometrical method to study the circular orbit of a particle moving in SSS spacetimes. On the other hand, we used the circular orbit to investigate the weak gravitational deflection of null and time-like particles. By applying the GB theorem to a finite area surrounded by four geodesics such as partial particle circular orbit, we obtain a new expression of deflection angle as follows
\begin{eqnarray}
\alpha&=&\iint_{D}\mathcal{K} dS+\phi_{RS}.\nn
\end{eqnarray}
This formula is suitable for studying the finite-distance effect of deflection angles, and also for weak gravitational deflection of particle in asymptotically non-flat spacetimes. In particular, we show that
{\begin{eqnarray}
 \left[\int\mathcal{K}\sqrt{det g}dr\right]_{r=r_{co}}=0,\nn
\end{eqnarray}}
where we ignore the constant of integration. It is very beautiful, and in practice, it can simplify the calculation of the deflection angle. In addition, we apply our proposed formula to calculate the weak gravitational deflections of light in a SSS spacetime in Weyl gravity and the deflection of massive in a Schwarzschild-like spacetime in bumblebee gravity, respectively, and the results are consistent with the previous literature. Finally, it will be our future work to extend present method to stationary spacetimes.

\acknowledgements
Z.L. thanks his family for taking care of him during such a difficult time.

\end{document}